\documentclass[prl,twocolumn,showpacs,amsbsy,floatfix,amsbsy,floatfix,superscriptaddress]{revtex4}
\usepackage{epsfig,color}
\usepackage[cp1250]{inputenc}
\usepackage{amsmath}
\usepackage{color}

\begin{document}

\title{Spin Orbit Mediated Manipulation of Heavy Hole Spin Qubit in Gated Semiconductor Nanodevices}

\author{P. Szumniak}
\affiliation{AGH University of Science and Technology, Faculty of
Physics and Applied Computer Science,\\
al. Mickiewicza 30, 30-059 Krak\'ow, Poland}
\affiliation{Departement Fysica, Universiteit Antwerpen, Groenenborgerlaan 171,
 B-2020 Antwerpen, Belgium}
\author{S. Bednarek}
\affiliation{AGH University of Science and Technology, Faculty of
Physics and Applied Computer Science,\\
al. Mickiewicza 30, 30-059 Krak\'ow, Poland}
\author{B. Partoens}
\affiliation{Departement Fysica, Universiteit Antwerpen, Groenenborgerlaan 171,
 B-2020 Antwerpen, Belgium}
\author{F. M. Peeters}
\affiliation{Departement Fysica, Universiteit Antwerpen, Groenenborgerlaan 171,
 B-2020 Antwerpen, Belgium}

\date{\today}

\begin{abstract}
A novel spintronic nanodevice is proposed that is capable to manipulate the single heavy hole spin state in a coherent manner. It can act as a single quantum logic gate. The heavy hole spin transformations are realized by transporting the hole around closed loops defined by metal gates deposited on top of the nanodevice. The device exploits Dresselhaus spin orbit interaction which translates the spatial motion of the hole into a rotation of the spin. The proposed quantum gate operates on sub nanosecond time scales and requires only the application of a weak static voltage which allows for addressing heavy hole spin qubit individually. Our results are supported by quantum mechanical time dependent calculations within the four band Luttinger Kohn model.
\end{abstract}


\maketitle

There is currently great interest in studying spin related phenomena in semiconductors. On the one hand there is novel fundamental physics at the nanoscale and on the other hand one expects applications in terms of spin based quantum information processing\cite{qd2,qd2a}.
Physical realization of quantum computers requires fulfillment of a number of challenging criteria\cite{qd1a}. A fragile quantum state has to be coherent for sufficient long time which usually requires its isolation from the environment. On the other hand it has to be externally manipulated. For these purposes, the electron spin in semiconductor quantum dots was suggested as a promising candidate\cite{qd1}.
There are a number of experiments in which the electron spin is initialized, manipulated, stored and read out\cite{qd3, qd4, qd4a, qd5, qd6, qd7, qf7a}. \\
\indent
Usually spin state manipulation requires the application of microwave radiation, radio-frequency electric fields as well as magnetic fields. These methods strongly limits the possibility to address spins qubits individually.\\
The first step towards selective control of individual single electron spins  was demonstrated in recent state of the art experiments\cite{qd8,qd29}. Electron spin manipulation was realized by means of electric fields which can be generated locally quite easily and indirectly via spin orbit interaction which couples charge and spin degrees of freedom. Electron spin control based on spin orbit effect was also proposed in some theoretical papers\cite{qd90,qd9a,qd9b,qd9, qd40}.\\
\indent
Unfortunately, in most semiconductor quantum dots the electron spin is exposed to hyperfine interaction with nuclear spins which are present in the host material. This interaction is then the main source of  electron spin decoherence in quantum dots putting a severe restriction on the possibility to realize a highly coherent electron spin qubit\cite{qd60,qd60a}.\\
%
There are several appealing ideas how to deal with this type of decoherence in quantum dot systems\cite{qd50}.
Very promising way to eliminate or reduce the contact hyperfine interaction with the nuclear spin lattice is to use the spin state of the valence holes - a missing electron in the valence band - as a carrier of quantum information instead of electrons. Holes are described by the p-orbitals that vanish at a nuclear site which strongly suppresses the hyperfine contact interaction. Thus one can expect longer coherence times for hole spin states\cite{qd28,qd291}. Some experiments seem to confirm this statement reporting long relaxation ($\sim$ms) and coherence ($\sim\mu$s) times \cite{qd23,qd26,qd88,qd55} for hole spins while others reported a very short hole spin dephasing time ($\sim$ns)\cite{qd52}. Recent theoretical investigations\cite{qd53,qd25} and experiments \cite{qd17} seem to resolve this mismatch of coherence time in different experiments suggesting that the absence of mixing between the heavy hole (HH) and the light hole(LH) state is crucial for a long hole spin coherence time.
Not only long coherence times but also the possibility to initialize the hole spin state even without a magnetic field\cite{qd26}, and the recent realization of a coherent control of a hole spin state in single and double coupled quantum dots\cite{qd18,qd54,qd55a} has promoted the hole as a very good candidate as carrier of quantum bit information.
There are also a few theoretical proposals how the HH spin state can be manipulated \cite{qd32, qd31, qd32aa}.\\
\indent
In this paper we demonstrate by using a four band HH-LH model that the motion of the valence hole in gated semiconductor nanostructures can induce the rotation of the HH spin in the presence of the Dresselhaus spin-orbit interaction (DSOI). Supported by these results we present an efficient scheme which can be used to realize any rotation of the HH spin and propose a nanodevice which acts as a quantum logic NOT gate on a HH spin qubit. The spin rotations are realized by transporting the hole along a closed loop which is defined by metal gates. Application of the multiband model allows us to study mixing between HH and LH states. We found that in the considered nanostructures the HH-LH mixing is negligible so we can expect long coherence times for a qubit stored in the HH spin state.\\
\indent
We consider a planar heterostructure covered by nanostructured metal gates. The system consist of a $10$ nm thick zinc-blende quantum well structure sandwiched between two 10 nm blocking barriers (Fig. 1) in which the single valence hole is confined.

\begin{figure}[ht!]
\epsfxsize=70mm \epsfbox[12 214 278 322]{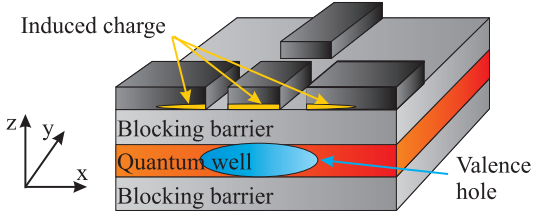}
\caption{Crossection of the nanodevice}
\end{figure}
The hole which forms a charge distribution (associated with its wave function) in this quantum well induces a response potential of the electron gas in the metallic gate which in turns leads to a self-focusing mechanism of the confined charged particle wave function \cite{qd11}. Thus interaction of the hole with the metal is a source of additional lateral confinement. As a result the hole is self-trapped under the metal in the form of a stable Gaussian like wave packet. It has the unique property for a quantum particle, that it reflects from a barrier or tunnels through it with 100$\%$ probability while conserving its shape, which is rather a characteristic of classical objects. This property can be used to transfer a charged particle in the form of a stable wave packet (soliton) between different locations within the nanodevice by applying static weak voltages to the electrodes only\cite{qd12}.\\
We use a system of coordinates in which the quantum well is oriented in the $z[001]$ (growth) direction and the hole can move only in the $x[100]-y[010]$ plane. We consider the two dimensional four band HH ($J_z=\pm 3/2$), LH ($J_z=\pm 1/2$) Hamiltonian:
\begin{equation}
    \hat{H}=\hat{H}_{LK}-|e|\phi(x,y,z_0)\hat{I}+\hat{H}_{BIA}^{2D}.
\end{equation}
The first term is the Luttinger-Kohn Hamiltonian\cite{qdLutt} describing  the kinetic energy of the 2D hole, which for zinc blende materials can be written in the effective mass approximation as:
\begin{equation}
\hat{H}_{LK}= \left( \begin{array}{cccc}
\hat{P}_{+} & 0 & \hat{R} & 0 \\
0 & \hat{P}_{-} & 0 & \hat{R} \\
\hat{R}^{\dag} & 0 & \hat{P}_{-} & 0  \\
0 & \hat{R}^{\dag} & 0 & \hat{P}_{+}  \end{array} \right),
\end{equation}
where $\hat{P}_{\pm}=\frac{\hbar^2}{2m_0}(\gamma_1\pm\gamma_2)(k_x^2+k_y^2)+E_0^\pm$ and $\hat{R}=\frac{\hbar^2}{2m_0}\sqrt{3}[\gamma_2(k_x^2-k_y^2)-2i\gamma_3k_xk_y]$. We denote $E_0^\pm=\frac{\hbar^2}{2m_0}(\gamma_1\mp 2\gamma_2)\langle k_z^2\rangle $ as the first subband energy in the $z$ direction with $\langle k_z\rangle ^2=\pi^2/d^2$, where $d$ is the quantum well height, $\gamma_1, \gamma_2, \gamma_3$ are the Luttinger parameters and $m_0$ is the free electron mass. Momentum operators are $k_q=-i\frac{\partial}{\partial q}$ where $q=x,y$. We use the representation where the projections of Bloch angular momentum on the $z$ axis are arranged in the following order $J_z=\frac{3}{2}, \frac{1}{2}, -\frac{1}{2}, -\frac{3}{2}$. Consistently with this convention the state vector can be written as
\begin{eqnarray}
\Psi(x,y,t)=(\psi_{HH}^\uparrow(x,y,t),\psi_{LH}^\uparrow(x,y,t),\\ \psi_{LH}^\downarrow(x,y,t),
\psi_{HH}^\downarrow(x,y,t))^T.
\end{eqnarray}
The electrostatic potential $\phi(x,y,z_0,t)$ which is "felt" by the hole is the source of the self trapping potential. Its origin is due to charges induced on the metal electrodes. The potential is found by solving the Poisson equation in a three dimensional computational box containing the entire nanodevice. 
The detailed method was described in Refs \cite{qd12, qd9}. Quantum calculations
\cite{qd13} indicate that this is a good approximation of the actual response potential of the electron gas. The $\hat{I}$ is the unit operator, $e$ is the elementary charge and $z_0$ is the center of the quantum well. The $\hat{H}_{BIA}$ term accounts for the DSOI 
\cite{qd14} which is caused by the lack of inversion symmetry of the crystal - a characteristic feature for zinc blende materials - and (including two main contributions) takes the following form for bulk\cite{qd39}
\begin{equation}
\hat{H}_{BIA}=-\beta_0\boldsymbol{k}\cdotp\boldsymbol{\Omega_J}-\beta  \boldsymbol{\Omega_k}\cdotp\boldsymbol{J},
\end{equation}
where $\boldsymbol{k}=(k_x, k_y, k_z)$ is the momentum vector and $\boldsymbol{J}=(J_x, J_y, J_z)$ is the vector of the $4\times 4$ spin $3/2$ matrices. The $x$ component of $\boldsymbol{\Omega_O}$ is the $\Omega_O^x=\{O_x,O_y^2-O_z^2\}$ and $\Omega_O^y,\Omega_O^z$ can be obtained by cyclic permutations, 
$\{{A},{B}\}=\frac{1}{2}({A}{B}+{B}{A})$ and the operator $O=k,J$.
Going from bulk to 2D systems and neglecting qubic $k$ terms the bulk DSOI can be directly transformed into:
\begin{eqnarray}
\hat{H}_{BIA}^{2D}&=&
\nonumber
\frac{\beta_0}{2}\left( \begin{array}{cccc}
0 & \sqrt{3}k_+ & 0 & 3k_- \\
\sqrt{3}k_- & 0 & -3k_+ & 0 \\
0 & -3k_- & 0 & \sqrt{3}k_+  \\
3k_+ & 0 & \sqrt{3}k_- & 0  \end{array} \right)\\
&+&
\nonumber
\frac{\beta \langle k_z^2\rangle}{2}\left( \begin{array}{cccc}
0 & \sqrt{3}k_+ & 0 & 0 \\
\sqrt{3}k_- & 0 & 2k_+ & 0 \\
0 & 2k_- & 0 & \sqrt{3}k_+  \\
0 & 0 & \sqrt{3}k_- & 0  \end{array} \right)\\
\end{eqnarray}
where $k_{\pm}=k_x\pm ik_y$ and values of $\beta_0$ and $\beta$ for different materials can be found in Refs \cite{qd39, qd39a}.\\
The time evolution of the system is described by the time dependent Schr\"{o}dinger equation which is solved numerically self-consistently with the Poisson equation. Due to the motion of the hole wave packet the Poisson equation has to be solved in every time step of the iteration procedure. The initial condition is the ground state of the hole confined under the metal due to the self focusing effect and is calculated by solving the stationary Schr\"{o}dinger equation $\hat{H}\Psi_0(x,y)=E\Psi_0(x,y)$.\\
\indent
Let us consider the heterostructure from Fig. 1 covered by the two electrodes $e_1$ and $e_2$ depicted in Figs 2$(d)$, 2$(d')$ respectively. In the initial state the hole is confined in the ground state under $e_1$ which can be achieved by applying a voltage $V_1=-0.3$ mV and $V_2=0$ to the electrodes $e_1$ and $e_2$, respectively. We assume that the hole is in the initial state $\Psi(x,y,t_0)=(
\psi_{HH}^\uparrow(x,y,t_0),0,0,0)^T$ \cite{comment1}. The hole is forced to move along the path under the electrode $e_2$ by changing the voltage configuration to $V_1=0$ and $V_2=-0.5$ mV. We plot the probability of finding the hole in the possible basis states $P_{J_z}(t)=\int|\psi_{J_z}(x,y,t)|^2dxdy$ in Figs 2$(c)$, 2$(c')$ where $J_z=3/2,1/2,-1/2, -3/2$. We observe that during the motion as well as in the ground state, the probability of finding the hole in the LH state is very small $(\sim 10^{-4})$. It means that the mixing between HH and LH states is negligible. 
\begin{figure}[ht!]
\epsfxsize=83mm \epsfbox[18 254 577 583]{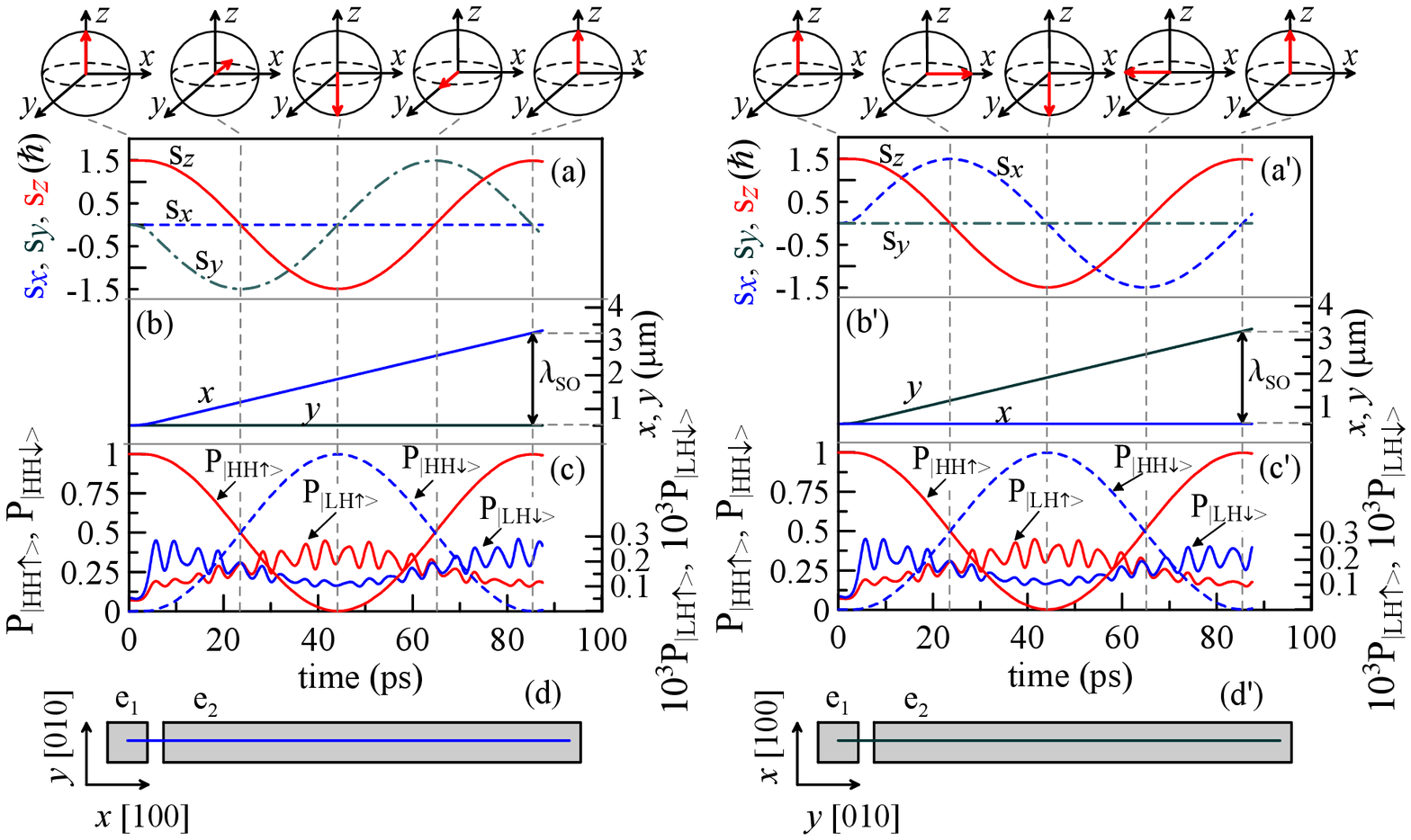}
\caption{Time evolution of the HH spin components (a), average position of the hole wave packet(b) and probability of occupying the following hole basis states(c): $|HH\uparrow\rangle$, $|HH\downarrow\rangle$, $|LH\uparrow\rangle$, $|LH\downarrow\rangle$, for a hole moving along the wire covered by the electrodes $e_1$, $e_2$ form figure (d). In figure (c) left (right) axis corresponds to probability of finding the hole in the HH (LH) spin states. Results for hole moving along the wire placed in y (d') direction are depicted in (a') (b') (c'). Above each plot there are Bloch spheres representing the qubit after each $\pi/2$ rotation.}
\end{figure}
Because of the fact that the hole is mainly composed of the HH state in the considered nanostuctures we can calculate expectation values of the HH pseudospin $1/2$ operator $\vec{s}=\langle\frac{3\hbar}{2}\vec{\sigma}\rangle_{\Psi_{HH}}$ for the HH state defined as $\Psi_{HH}(x,y,t)=(\psi_{HH}^\uparrow(x,y,t), \psi_{HH}^\downarrow(x,y,t))^{T}$  where $\vec{\sigma}$ are the Pauli spin $1/2$ matrices. For a hole occupying the HH band only the expectation values of total angular momentum $J=3/2$ matrices are $\langle J_x\rangle=\langle J_y\rangle=0$ and $\langle J_z\rangle=s_z$. The time dependence of the HH average spin components are given in Figs 2$(a)$, 2$(a')$.
During the motion of a hole along the $x$ $(y)$ axis the $s_x$ $(s_y)$ spin component is preserved and $s_y$ $(s_x)$, $s_z$ components oscillate: the HH spin is rotated around the axis parallel to the direction of motion.
This behavior can be understood by analyzing an approximated $\hat{H}_{BIA}$ Hamiltonian for the HH band only\cite{qd30} 
\begin{equation}
    \hat{H}_{BIA}^{HH}=-\frac{\tilde{\beta}}{\hbar^3}[(p_{x}^{3}+p_{x}p_{y}^{2} )\sigma_{x}+  (p_{y}^{3}+p_{x}^{2}p_{y} )\sigma_{y}].
\end{equation}
For quantum wires placed along the $q$ direction, the above Hamiltonian can be approximated by $\hat{H}_{BIA,q}^{HH}=-\frac{\tilde{\beta}}{\hbar^3}(p_{q}^{3}+p_{q}\langle p_{q_{\perp}}^{2} \rangle )\sigma_{q}$, where $q=x,y$, $q_{\perp}$ axis is perpendicular to $q$ and the $\tilde{\beta}$ is an effective DSOI coupling strength given in \cite{qd30}. \\
From the fact that the momentum operators $p_q$ and $p_q^3$ are multiplied by the HH spin operator $\sigma_q$, one can expect that the hole motion with $p_q$ momentum will generate a spin rotation around the $q$ axis according to the time evolution operator $\hat{U}_q(t)=e^{-i\hat{H}_{BIA,q}^{HH}t/\hbar}$.\\
\indent
After traveling a certain distance $\lambda(t)$, the  spin is rotated by the angle $\phi(t) =2\pi\frac{\lambda(t)}{\lambda_{SO}}$. Thus one can say that a unitary operation was performed on the HH spin state. One can derive the corresponding unitary spin rotation operator for a hole moving in the wire placed along the $x$ axis:
\begin{equation}
    \hat{R}_x(\phi)=\frac{1}{\sqrt{2}}\left( \begin{array}{cc}
i\sqrt{1+\cos(\phi)} & \frac{\sin(\phi)}{\sqrt{1+\cos(\phi)}} \\
\frac{\sin(\phi)}{\sqrt{1+\cos(\phi)}} & i\sqrt{1+\cos(\phi)}\end{array} \right)
\end{equation}
and for a hole moving in the wire which is placed along the $y$ direction:
\begin{equation}
    \hat{R}_y(\phi)=\frac{1}{\sqrt{2}}\left( \begin{array}{cc}
\sqrt{1+\cos(\phi)} & -\frac{\sin(\phi)}{\sqrt{1+\cos(\phi)}} \\
\frac{\sin(\phi)}{\sqrt{1+\cos(\phi)}} & \sqrt{1+\cos(\phi)}\end{array} \right).
\end{equation}
\indent
The hole restores its initial spin after passing the distance $\lambda_{SO}$ which depends on the DSOI coupling strengths and the effective mass(Luttinger parameters).
\indent
The presented results are obtained for a GaAs quantum well, but
we also performed calculations for other materials ZnSe, and CdTe and estimated the $\lambda_{SO}$ length: $\lambda_{SO}^{GaAs}=2.75$ $\mu$m , $\lambda_{SO}^{ZnSe}=0.6$ $\mu$m, $\lambda_{SO}^{CdTe}=0.5$ $\mu$m \cite{parameters}.\\
It's worth to mention that the "on demand" single electron transport on such distances ($\mu$m) and even much larger was recently realized experimentally using surface acoustic waves \cite{100,101}.\\
\indent
Taking advantage of the fact that the hole motion generates HH spin rotations one can design a gated semiconductor nonodevice which will act on the HH spin qubit as a quantum gate. We propose a nanodevice covered by the system of electrodes from Fig. 3 which act as a quantum NOT gate.

\begin{figure}[ht!]
\epsfxsize=83mm \epsfbox[14 120 577 718]{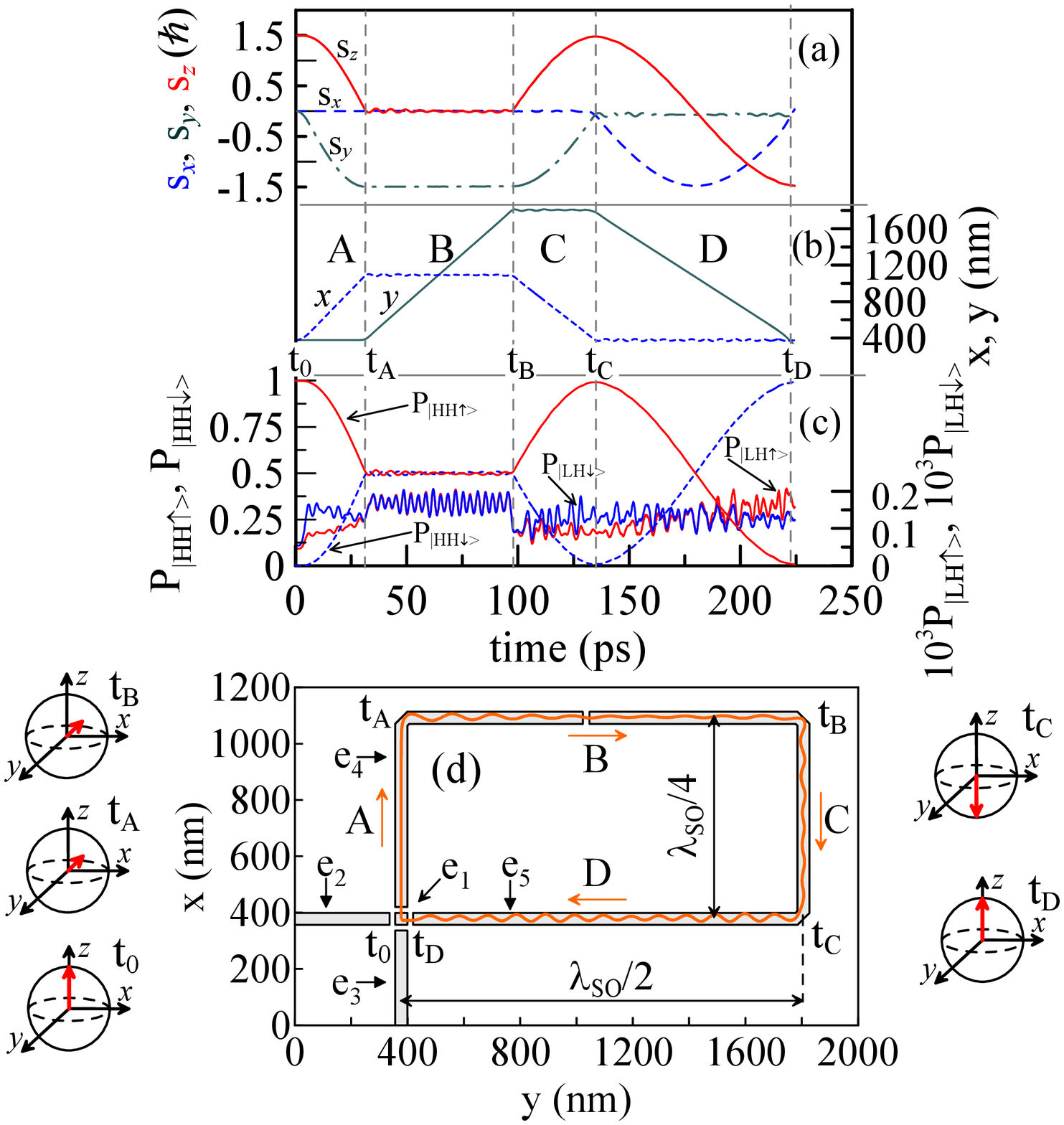}
\caption{ Same as Fig 2. but for the quantum NOT gate which is covered by the system of electrodes $e_1-e_5$ presented in (d). In figure (d) solid orange line represents the hole trajectory (orange arrow represents direction of motion of the hole). Hole initially confined under $e_1$ goes to the +x direction passing A, B, C, and D segments of the loop. Vector state is depicted on the Bloch spheres in the subsequent moments of time: $t_0-t_D$.}
\end{figure}
The hole whose spin we want to transform is initially confined under electrode $e_1$ where a constant $V_1=-0.2$ mV voltage is applied while the voltage on the other electrodes is set to zero. Let us assume that the hole is in the ground state with its initial spin state prepared to the HH spin up state. By changing the voltage applied to $e_1$ to $V_1=0$ and increasing the voltage on $e_4$ to $V_4=-0.5$ mV the hole starts to move in the $+x$ direction under electrode $e_4$. After passing a $\lambda_{SO}/4$ distance of segment A the HH spin is rotated around the $x$ axis by an angle $\pi/2$ and the $\hat{R}_x(\pi/2)$ operation is performed. At the end of segment A the hole wave packet turns right and starts to move parallel to the $y$ axis. During the reflection the hole wave packet does not scatter due to the self-focusing effect. The hole goes under electrode $e_5$ whose voltage was in the meantime set to the voltage of the $e_4$ electrode. The hole passes the B segment whose length is $\lambda_{SO}/2$ performing the $\hat{R}_y(\pi)$ operation and turns right. Then the hole moves in $-x$ and $-y$ directions performing the $\hat{R}_x(-\pi/2)$ and $\hat{R}_y(-\pi)$ operation. Finally the hole returns to its initial position under the $e_1$ electrode where it is captured by applying the $V_1=-0.7$ mV voltage.
After passing the whole loop a set of HH spin transformations is performed resulting in a NOT gate operation $\hat{U}_{NOT}=\hat{R}_y(-\pi)\hat{R}_x(-\pi/2)\hat{R}_y(\pi)\hat{R}_x(\pi/2)=i\sigma_x$.
Since the hole after completing the set of transformations returns to its initial position, the gate operation is performed on the HH spin exclusively, not on the spatial part of the wave function. 
Size of the gate depends only on the $\lambda_{SO}$ length for the considered material.\\
\indent
The gate operation time for GaAs and applied starting voltage configuration is $t_{NOT}^{GaAs}\approx 220$ ps. As the time is proportional to $\lambda_{SO}$ the gate operation time for other materials is significantly improved reaching $t_{NOT}^{CdTe}\approx 40$ ps and $t_{NOT}^{ZnSe}\approx 50$ ps.\\
\indent
In conclusion, we showed that the motion of the hole in gated semiconductor heterostructures can induce a coherent rotation of the HH spin where the DSOI is the mediator of this process. An important result is that during the motion in the presence of the DSOI the mixing between HH and LH states is negligible from which we can expect that the proposed HH spin qubit should be robust to decoherence coming from the interaction with the nuclear spins. We proposed a quantum NOT gate which operates in sub nanoseconds and it is controlled only by means of small static local electric fields generated by the top gates. It allows to address the HH spin qubit individually making our proposal scalable.
{\it Acknowledgements.}
This work was supported by the Grant No. N N202 128337 from the Ministry of Science and Higher Education, as well as by the "Krakow Interdisciplinary PhD-Project in Nanoscience and Advances Nanostructures" operated within Foundation for Polish Science MPD Programme co-financed by European Regional Development Fund.


\begin{thebibliography}{00}
\bibitem{comment1} In fact due lateral confinement ground state of the hole in induced quantum dot has some substantial amount $(<0.1\%)$ of LH states Fig. 2c.
\end{thebibliography}
\end{document}